\documentclass[11pt]{article}
\usepackage{graphicx}
\usepackage{amsmath,amsfonts,amssymb}
\usepackage{color}
\usepackage{bbold}
\def\sloppy{\tolerance=100000\hfuzz=\maxdimen\vfuzz=\maxdimen}
\allowdisplaybreaks
\def \beq  {\begin{equation}}
\def \eeq  {\end{equation}}
\def \beqar {\begin{eqnarray}}
\def \eeqar {\end{eqnarray}}
\def\bsp{\beq\begin{split}}
\def\esp{\end{split}\eeq}
\mathchardef\mhyphen="2D

\def\la {{\langle}}
\def\ra {{\rangle}}

\def\vf {{\varphi}}

\def\Tr {{\rm Tr}}
\def \tr {{\rm tr}}

\def\bD {\bar{D}}
\def\bA {\bar{A}}
\def\bE {\bar{E}}

\def\del {\partial}
\def\bdel{\bar{\partial}}
\def\a {\alpha}

\def\e {\epsilon}

\def\A {{\cal A}}
\def\C {{\cal C}}

\def\E {{\cal E}}

\def\G {{\cal G}}
\def\H {{\cal H}}

\def\O {{\cal O}}

\def\S {{\cal S}}

\def\half{\textstyle{1\over 2}}

\begin{document}
\begin{titlepage}
\null\vspace{-62pt} \pagestyle{empty}
\begin{center}
\rightline{CCNY-HEP-16/8}
\rightline{December 2016}
\vspace{1truein} {\Large\bfseries
Gauge-invariant Variables and Entanglement Entropy}\\
\vskip .15in
{\Large\bfseries ~}\\
\vskip .1in
{\Large\bfseries ~}\\
{\large\sc Abhishek Agarwal$^a$},
{\large\sc Dimitra Karabali$^b$} and
 {\large\sc V.P. Nair$^c$}\\
\vskip .2in
{\itshape $^a$\,Physical Review Letters\\
American Physical Society\\ 
Ridge, NY 11961}\\
\vskip .1in
{\itshape $^b$Department of Physics and Astronomy\\
Lehman College of the CUNY\\
Bronx, NY 10468}\\
\vskip .1in
{\itshape $^c$Physics Department\\
City College of the CUNY\\
New York, NY 10031}\\
\vskip .1in
\begin{tabular}{r l}
E-mail:&{\fontfamily{cmtt}\fontsize{11pt}{15pt}\selectfont abhishek@aps.org}\\
&{\fontfamily{cmtt}\fontsize{11pt}{15pt}\selectfont dimitra.karabali@lehman.cuny.edu}\\
&{\fontfamily{cmtt}\fontsize{11pt}{15pt}\selectfont vpnair@ccny.cuny.edu}
\end{tabular}

\vspace{.5in}
\centerline{\large\bf Abstract}
\end{center}
The entanglement entropy (EE) of gauge theories in three spacetime dimensions is analyzed using
manifestly gauge-invariant variables defined directly in the continuum. Specifically, we focus on
the Maxwell, Maxwell-Chern-Simons (MCS),  and nonabelian Yang-Mills theories.
Special attention is paid to the analysis of edge modes and their contribution to EE. The contact term is derived without invoking the replica method and its physical origin is traced to the phase space volume measure for the edge modes. The topological contribution 
to the EE for the MCS case is calculated.
For all the abelian cases, the EE presented in this paper agrees with 
known results in the literature.  
The EE  for the nonabelian theory is
computed in a gauge-invariant gaussian approximation,
which incoprorates the dynamically generated mass gap.
 A formulation of the contact term for the nonabelian case is also presented.
\end{titlepage}

\pagestyle{plain} \setcounter{page}{2}
\setcounter{footnote}{0}
\setcounter{figure}{0}
\renewcommand\thefootnote{\mbox{\arabic{footnote}}}
\section{Introduction and Summary}
In this paper we formulate a gauge-invariant set up,
defined directly in the continuum,
 for computing the entropy of vacuum entanglement (EE) for gauge theories in three spacetime dimensions.
 We apply this approach to analyze the EE for 
the abelian Maxwell,  Maxwell-Chern-Simons (MCS) and nonabelian pure Yang-Mills theories.
Our approach allows for the identification and clarification of the physical origin
of the Kabat contact term for all these cases
without invoking the replica trick and the ensuing conical partition function. 
The calculational framework is set up in such a way
that it can be extended to more involved geometries than have been studied in the past. 

Quantifying the information content of gauge theory vacua has recently emerged as an interesting computational as well as a conceptual challenge. While it can be argued - in analogy with finite dimensional quantum mechanical models  - that EE, defined as the von Neumann entropy of the reduced density matrix $\rho_{red}$ is a reasonable measure of the ground state entanglement of any field theory, its definition and computation are far from obvious in a gauge theory.  
In a typical field theory without gauge symmetries (e.g a free scalar theory),
even though the wave functionals do display entanglement,
they, as well as observables, are expressed in terms of local fields.
Thus, there is no conceptual  obstruction to integrating out degrees of freedom for the ``inside" region (or the ``outside"
region) to obtain a reduced density matrix for the complementary region.
This can be done just as one would in a system of coupled oscillators, and this was indeed the approach taken in the early computations of EE for scalar field theories \cite{early}. 
By contrast, the obvious physical degrees of freedom in a gauge theory
are the nonlocal Wilson loops which can prevent a clean separation of the Hilbert space into ``inside'' and ``outside'' regions. Further, \cite{early} used an explicit position space representation of  the vacuum wave functional $\Psi$ for the scalar theory. Our explicit knowledge of gauge invariant $\Psi$'s is limited at best for gauge theories.

Several different approaches to circumvent these issues have recently been explored. One could conceive of defining and computing EE on a lattice \cite{lattice}. Though the lattice allows one to compute using quasi-local link and plaquette variables, the constraint of gauge invariance
again impacts on separating variables into different
regions.
This leads to several prescriptions for ``cutting'' the lattice into ``inside'' and ``outside'' regions and they may lead to inequivalent expressions for EE. (See \cite{lattice3d} for a computation explicitly devoted to these issues for the Maxwell theory in $D=2+1$.) Further, a topological term
such as the Chern-Simons term 
is difficult to realize on the lattice.  
An alternative method, the replica trick,  allows one to work directly in the continuum
by formally expressing the wave functional as a Euclidean path integral with specific boundary conditions at the initial and final time slices \cite{scalar, reviews}.\footnote{ Some subtleties in the use of replica trick for EE for gauge theories have been discussed in \cite{related-cone}.} 
The $\alpha$-th power of the reduced density matrix
$\rho_{red}$ is formally expressed as a path integral over a cone using the replica method. 
However, a rigorous proof of equivalence of the entropy computed from the cone partition function ($S_{cone}$) with EE (as computed from an explicit construction of $\Psi$) is  lacking for gauge theories.\footnote{However $S_{cone}$ is interesting in its own right as it computes the one-loop correction to the Bekenstein-Hawking entropy of a black hole.}
In a beautiful paper, Kabat \cite{kabat} computed $S_{cone}$ for Maxwell fields which have $D-2$ physical degrees of freedom. The computation in \cite{kabat} showed that $S_{cone}$ differed from the EE of $D-2$ scalar fields by an additional negative contribution; the so called contact term. The contact term and its possible physical interpretation has been explored recently \cite{DW,CH}, however it is is not completely clear if it arises in computations of EE entirely within a
Hamiltonian formulation \cite{CH}. A natural question to ask is whether the contact term is an artifact of the replica trick and whether it can be ascribed a physical significance without recourse to the replica method. In this work we address this question specifically in three spacetime dimensions and find that the contact term arises naturally from the phase space
 integration measure.
 The result can also be generalized - at least formally -  to the nonablelian case and to more involved geometries than are analytically accessible by the replica procedure.

For purposes of computing ``half-space'' entanglement between two regions ${\rm I}$ and ${\rm II}$, our strategy is to decompose the Hamiltonian  and invariant degrees of freedom of the theory defined in ${\rm I} \bigcup {\rm II}$  such that we have a manifestly gauge invariant  formulation of the theory individually in ${\rm I}$ and ${\rm II}$. It is crucial to the decomposition  to note that the gauge transformations that do not go to  identity at spatial boundaries must be regarded as physical degrees of freedom; namely the edge modes of gauge theories \cite{bal}. Such edge modes would be present at  the interface between ${\rm I}$ and ${\rm II}$ were it to be a real boundary. The process of 
entangling ${\rm I}$ and ${\rm II}$ involves integrating the edge modes out of the physical Hilbert space, rather than just setting them to zero.
This, in turn, induces a degeneracy factor in the reduced density matrix. Formally, this degeneracy is captured by the measure of the phase space path integral and we show that it accounts precisely for the contact term. 
This construction generalizes to the MCS and, to some extent, to nonabelian cases. 
 In the case of the MCS theory, we also show that an additional contribution from the edge modes arises when the entangling surface has a nontrivial topology. Specifically for the case of a circular interface, the zero modes of the edge modes produce an additional topological contribution,
 which is identical to what is obtained for the pure Chern-Simons theory.
 
 To briefly recapitulate, the key new results of this paper are
 the following.
We formulate the computation of the EE in 2+1 dimensional gauge theories
in terms of gauge-invariant variables and apply this to
the abelian Maxwell, the Maxwell-Chern-Simons and (approximately)
to pure Yang-Mills theories.
The origin of the contact term is identified as arising from
integration of the edge degrees of freedom on the interface, rather than 
factoring them out as gauge degrees of freedom.
The contact term is shown to be related to the 
interface term which arises in the Mayer-Vietoris type decomposition
of the determinant of the Laplacian on a Riemannian manifold, namely
the BFK gluing formula \cite{BFK}. This is also the term which is
important in capturing the diffractive contributions
in the Casimir effect \cite{cas1}.
For the Maxwell-Chern-Simons theory, with a circular interface separating the two
regions
of interest, we show that there is also a topological contribution which is identical to what is obtained
for the Chern-Simons theory modulo the usual
regularization-dependent terms.
For the nonabelian theory, we find a tractable regime where the EE for pure Yang-Mills theory can be computed, albeit approximately.
It is possible to include the effect of the nonperturbative mass
gap and also obtain an approximate expression for the contact term.

This paper is organized as follows. In section 2,
we collate and review some basic results for half-space entanglement for the massive scalar field in three spacetime dimensions, which we need to call on for later discussions.
Section 3 is devoted to the Maxwell theory in terms of gauge-invariant variables,
keeping track of
the edge modes. The phase space measure of these modes is shown to account 
for the contact term after the modes are integrated out of the physical spectrum. 
In section 4, we do a similar analysis
for the MCS theory. The contact term contribution remains the same as in the pure Maxwell case.
The bulk EE is now given by that of a massive scalar. 
We also study the MCS theory with a circular entangling surface in which case we recover the well known topological term known to be present in the case of pure Chern-Simons theory.  
In section 5, we discuss
the nonabelian pure-glue theory. 
Of course, this theory is not exactly solvable, and the computation of the EE,
 perforce, can only be approximate.
We use the formulation due to \cite{KKN} to obtain a gauge invariant truncation of the vacuum wave-functional of the gauge theory. This tractable regime
 allows us to work with a Gaussian wave functional which, nevertheless, retains information about the nonperturbative mass-gap of this theory. Specifically, we find that the mass-gap leads to a finite (and hence universal) term in the EE which is proportional to the area of the entangling surface. Interestingly, its contribution is negative, leading to a reduction in entanglement and provides a direct potential link between the IR properties of the gauge theory and EE. We also provide a general, although somewhat formal,
expression for the contact term for the nonabelian theory. 
In the final section we comment on some conceptual connections between our approach and the replica-trick methods.
There are two appendixes, one on a technical point on eliminating a certain field component
and the other on the topological contribution to the EE for the MCS theory.

\section{The entanglement entropy of a massive scalar field}

Since many of the cases we discuss will utilize the EE of a massive scalar field,
we start by briefly recalling how such a computation is carried out; for more details, see
\cite{reviews}.
With a standard action given by
\beq
\S= {1\over 2} \int d^3x \left[ (\del \phi )^2 - m^2 \phi^2 \right]
\label{ms1}
\eeq
the ground state (or vacuum) wave function is given by
\beq
\Psi [\vf ] = {\mathcal N} \, \exp \left( - {1\over 2} \int d^2x\,d^2y~ \vf (x)\, ( \sqrt{m^2 - \nabla^2})_{x,y}
\, \vf (y) \right)
\label{ms2}
\eeq
where $\vf$ denotes the value of the field at a given time-slice, say, at $t =0$.
We can now apply
the replica trick to the wave function.
The first step involves giving  (\ref{ms2}) a path integral representation as \cite{reviews}
\beq
\Psi[\vf ] = \int [\mathcal{D} \phi '] \delta (\phi'(x_i, t = 0) - \vf) ~e^{-\int\left[ \frac{1}{2}\phi'(p^2+m^2)\phi' \right]}\label{psiphi}
\eeq
where we integrate over all fields from $t = -\infty$ to $t =0$.
Here we are using the Euclidean action, $t$ for this expression being the Euclidean time variable.
We simply regard the above as a mathematical representation of the wave functional without ascribing any fundamental meaning to the three-dimensional action appearing in the functional integral on the 
right hand side.
Repeating this construction $\alpha $ times and integrating the degrees of freedom on the negative $x$-axis (out of the density matrix constructed from the wave function)
reduces the EE computation to the evaluation of  a functional integral for a massive scalar field 
on a cone with deficit angle $2\pi(1-\alpha)$ \cite{reviews}. 
More precisely, the EE can be defined as
\beq
{S}_E = (\a \partial_\a - 1)W_\a.\label{eff-s}
\eeq
where, $W_\a$ is the effective action on the cone, defined in terms of the heat kernel $K(s)
= e^{- s(p^2 + m^2)}$ as
\beq
W_\a = -\frac{1}{2}\int_{\e^2} ^\infty \tr K(s)
\eeq
Using known techniques, we evaluate the conical partition function and express 
the entropy as
\beq
S_E(m) = \frac{2\pi {\mathbb A}}{6}\int_{\e^2} ^\infty \frac{ds}{(4\pi s)^{d/2}}e^{-m^2s}
\eeq
where ${\mathbb A},\e$ are the entangling area and UV cutoff respectively. This expression is written for 
any spacetime dimension $d$; more 
concretely, for $d=3$, we get
\beq
S_E(m) = \frac{{\mathbb A}}{12\sqrt{\pi}}\left(\frac{e^{-\e^2m^2}}{\e} -m\,\sqrt{\pi}\, \mathrm{erfc}(\e m)\,\right)\label{d3s}
\eeq
In the above expression, $\mathrm{erfc}(x)$ is the 
complementary error function; $1-\mathrm{erf}(x)$.
It is instructive to expand (\ref{d3s}) in powers of the UV cutoff $\e$ as
\beq
S_E(m) = \frac{{\mathbb A}}{12\sqrt{\pi}}\left(\frac{1}{\e} -m\sqrt{\pi} + \mathcal{O}(\e)\right) \label{sm}
\eeq
The leading term here is the EE of a massless scalar field, which is proportional to the
area of the entangling surface and is cut-off dependent. 
(In 2+1 dimensions, ${\mathbb A}$ is the length of the entangling surface.)
There is also a finite term proportional
to the mass. This may be unambiguously extracted by taking the $\epsilon \rightarrow 0$ limit
of
$S_E(m) - S_E (0)$. Notice also that
the mass correction tends to decrease the entanglement entropy.
Since correlators are of short range ($\sim 1/m$) in a massive
theory, we may expect the entanglement to be reduced, in agreement with
(\ref{sm}).

We now turn to the gauge theories.

\section{The Maxwell theory}

Our approach, as mentioned in the introduction, is to cast the gauge theory entirely in terms of
gauge-invariant variables and use it to calculate the entanglement entropy. 
The arguments presented, we believe, will help to clarify the role of the Kabat contact term. 
In three spacetime dimensions, the gauge field has one physical polarization, 
so that part of the entropy
for the amount of entanglement encoded in the quantum vacuum wave function
is identical to
the contribution of a single massless scalar.\footnote{The last reference in \cite{lattice} refers to this part of EE as the 'extractable contribution' where it was shown to count the number of correlated Bell pairs.}
However, the elimination of the gauge degrees of freedom and
the factorization of the reduced phase volume into two regions can lead to a second contribution.
We show that this extra contribution is indeed the contact term. It is related to the 
surface term in the BFK gluing formula for determinants of the Laplace-type operators
\cite{BFK}.
This elucidates the nature of the contact term
without invoking the replica trick and
without referencing conical singularities.

Since we will use a Hamiltonian framework, we start with the gauge $A_0 = 0$. 
In two dimensions, the
spatial components $A_i$ and the electric field $E_i$ have the general parametrization
\beq
A_i = \del_i \theta + \epsilon_{ij} \del_j \phi, \hskip .3in
E_i = {\dot A_i} = \del_i {\sigma} + \epsilon_{ij} \del_j {\Pi}
\label{1}
\eeq
(Since $E_i = {\dot A}_i$, $\sigma$ and $\Pi$ are ${\dot \theta}$, ${\dot \phi}$
in a Lagrangian description. But here we want to carry out a Hamiltonian
analysis, so these are independent phase space variables.)
Our strategy is to set up the 
theory in two regions I and II, with I $\cup$ II being the spatial manifold.
We then consider the theory defined on the full space
${\rm I} \cup {\rm II}$ but write it in terms of variables appropriate to
regions ${\rm I}$ and ${\rm II}$. It is then easy to see
that integrating over the variables in region ${\rm II}$
does not give the theory which was {\it a priori} defined
in ${\rm I}$. This discrepancy is due to entanglement.
Alternatively, we can put together the theories defined in each region to
obtain the full space theory via suitable matching conditions.
This will also bring out the entanglement and the contact term.

We begin the analysis starting with region I; the situation for II will be similar.
Since degrees of freedom on the boundaries will be important, we start with a
decomposition of the fields given by
\beqar
\theta_{\rm I} (x) &=& {\tilde \theta}_{\rm I}(x) + \oint_{\del {\rm I}} \theta_{0{\rm I}} (y)\,
n \cdot \del G (y,x)_{\rm I}\nonumber\\
\sigma_{\rm I} (x) &=& {\tilde \sigma}_{\rm I} (x) + \oint_{\del {\rm I}} \sigma_{0{\rm I}} (y)\,
n \cdot \del G (y,x)_{\rm I}
\label{new1}
\eeqar
Here the tilde-fields all obey Dirichlet boundary conditions, vanishing on $\del$I.
The boundary values are explicitly shown as the fields with a subscript $0$.
They are continued into the interior such that they obey the Laplace equation,
i.e.,
\beq
\nabla^2_x \, \oint_{\del {\rm I}} \theta_{0{\rm I}} (y)\,
n \cdot G (y,x)_{\rm I} = 0, \hskip .2in {\rm etc.}
\label{new2}
\eeq
The Green's function $G(y, x)$ obeys Dirichlet conditions on the boundary.
The decomposition of the fields as in (\ref{new1}) follows from Green's theorem.
We will also do a similar decomposition for the field $\Pi$,
\beq
\Pi_{\rm I} (x) = {\tilde \Pi}_{\rm I} (x) + \oint_{\del {\rm I}} \Pi_{0{\rm I}} (y)\,
n \cdot \del G (y,x)_{\rm I}
\label{new2a}
\eeq
A similar separation of modes for $\phi$ will emerge from the simplifications which follow.

The needed ingredients for the analysis are the canonical structure and 
the Hamiltonian expressed in terms of 
this parametrization (\ref{new1}), (\ref{new2a}).
The canonical one-form is given by
\beqar
\A&=& \int E_i \, \delta A_i\nonumber\\
&=&\int_{\rm I} \left[ (-\nabla^2 {\tilde \sigma}_{\rm I})\, \delta {\tilde \theta}_{\rm I}
+ {\tilde \Pi}_{\rm I} \, \delta B_{\rm I} \right]
+ \oint \Bigl(\sigma_{0{\rm I}} (y) M(y,x)_{\rm I}   + {\del}_\tau \Pi_{0{\rm I} } \Bigr) \delta \theta_{0{\rm I} } (x) \nonumber\\
&&\hskip .2in
+ \oint \Bigl(\Pi_{0{\rm I} } (y) M(y,x)_{\rm I}   - {\del}_\tau \sigma_{0{\rm I} } \Bigr) \delta \phi_{0{\rm I} } (x) \nonumber\\
&=&\int_{\rm I} \left[ (-\nabla^2 {\tilde \sigma}_{\rm I})\, \delta {\tilde \theta}_{\rm I}
+ {\tilde \Pi}_{\rm I} \, \delta B_{\rm I} \right]
+ \oint \E_{{\rm I}}\, \delta \theta_{0{\rm I} } (x) 
+ \oint Q_{{\rm I}}\, \delta \phi_{0{\rm I} } (x) 
\label{new3}
\eeqar
where $B = -\nabla^2 \phi$ is the magnetic field,  ${\del_{\tau}} = n_i \epsilon_{ij} \del_j$ is the tangential derivative on the boundary, and
\beqar
M(x, y)_{\rm I}  &=&  n\cdot \del_x \,n\cdot \del_y G(x,y)_{\rm I} |_{ x,~y ~on~ \del {\rm I}} \label{M}\\
\E_{{\rm I}}(x)&=& \oint_y \sigma_{0{\rm I}} (y) M(y,x)_{\rm I}   + {\del_{\tau}} \Pi_{0{\rm I}}(x) \nonumber\\
Q_{{\rm I}}(x) &=& \oint_y \Pi_{0{\rm I} } (y) M(y,x)_{\rm I}   - {\del_{\tau}} \sigma_{0{\rm I} } (x)
\eeqar
Notice that $\E_{{\rm I}}$ and $-Q_{{\rm I}}$ are related to the normal and tangential components of the electric field on the boundary,
although not exactly equal to them.
If we consider a large rectangular volume divided into two regions with a flat interface,
then $M = \sqrt{q^2}$ where $q$ is the momentum variable 
(wave vector in a Fourier decomposition of the fields) along the boundary
and we have the identity 
\beq
\del_x \del_y M^{-1} (x, y) =  M (x,y)
\label{new4}
\eeq
In this case, it is easy to see that we have a constraint $\C = \del_y \oint \E_{{\rm I}}(x) M^{-1}(x,y) + Q_ {{\rm I}}(x)= 0$.
This is a first class constraint and so we can enforce it choosing a conjugate
constraint $\phi_0 \approx 0$.
The canonical one-form thus reduces to 
\beq
\A = \int_{\rm I} \left[ (-\nabla^2 {\tilde \sigma}_{\rm I})\, \delta {\tilde \theta}_{\rm I}
+ {\tilde \Pi}_{\rm I} \, \delta B_{\rm I} \right]
+ \oint \E_{ {\rm I}}\, \delta \theta_{0{\rm I} }  
\label{new5}
\eeq
The essence of the constraint on $Q$ can be understood as follows. The definition of
$\phi$ via $B = - \nabla^2 \phi$ allows some freedom, a new ``gauge type " redundancy, since
$\phi$ and $\phi + f$ give the same $B$, if $\nabla^2 f =0$.
Such an $f$ is entirely determined by the boundary value as
in
\beq
f (x) = \oint f_0 (y) \, n\cdot \del_y G (y,x)
\label{new6}
\eeq
It is then possible to choose $f_0$ to obtain $\phi =0 $ on $\del$I, 
for the same magnetic field $B$. This freedom is reflected in the constraint and in 
our ability to choose the conjugate constraint $\phi_0 \approx 0$.

The canonical structure (\ref{new5}) shows that the phase volume is given by
\beq
d\mu_{\rm I} = [d{\tilde \sigma} d {\tilde \theta}]_{\rm I}\,
 [d \E\, d\theta_0]_{\rm I} \, [ d{\tilde \Pi} d B]_{\rm I} \,  \det (-\nabla^2)_{\rm I}
 \label{new7}
 \eeq
The determinant is to be calculated with Dirichlet conditions on the modes.
The Hamiltonian can be simplified in a similar way to obtain
\beqar
\H &=&\int  {1\over 2}\ \left[  (\nabla {\tilde \sigma}_{\rm I})^2 
+ ( \nabla{\tilde \Pi}_{\rm I} )^2 + B_{\rm I}^2 \right] + {1\over 2} \oint \E_{ {\rm I}}(x)
M^{-1}(x, y)_{\rm I} \,\E_{ {\rm I}} (y) \nonumber\\
&&\hskip .2in + {1\over 2} \oint \Pi_{0{\rm I}} (x)( M (x,y)- \del_x \del_y M^{-1}(x,y) ) \, \Pi_{0{\rm I}}(y)
\label{new8}
\eeqar
The last term is actually zero for the flat interface in infinite volume 
(which is the case we will continue to discuss), due to
(\ref{new4}). It is also zero for the interface being a circle.
We display it here to show how there can be extra terms for interfaces with
curvatures or for more general partitioning of the full space.

It is also useful to write down the phase space path integral since it provides a 
succinct way to capture
the effects of both the Hamiltonian (and hence the wave function) and the integration measure.
For this, we first recall that, for a theory with first class constraints
$\Phi^\alpha \approx 0$, and corresponding gauge fixing constraints $\chi^\beta \approx 0$,
the phase space path integral is given by
\beq
Z = \int d\mu\, \prod_\alpha \delta ( \Phi^\alpha ) \prod_\beta \delta ( \chi^\beta )
~ \det \{\Phi^\alpha, \chi^\beta\}~
e^{i \S}
\label{new9}
\eeq
where $d\mu$ is the phase volume (for the full phase space before reduction by constraints)
and $ \{\Phi^\alpha, \chi^\beta\}$ denotes the Poisson bracket (computed with the full
canonical structure).
For the Maxwell theory, in the Coulomb gauge with $\Phi \equiv \nabla \cdot E$ and
$\chi\equiv \nabla \cdot A $, this gives
\beq
Z = \int d\mu~ \delta (\nabla \cdot E)~\delta(\nabla \cdot A)~\det (-\nabla^2) ~e^{i \S}\label{coulz}
\eeq
where the action is given, for ${\rm I}$, by $\A$ and $\H$ as
\beq
\S_{\rm I} = \int_{\rm I} \left[ (-\nabla^2 {\tilde \sigma}_{\rm I})\, {\dot {\tilde \theta}}_{\rm I}
+ {\tilde \Pi}_{\rm I} \, {\dot B}_{\rm I} \right]
+ \oint \E_{{\rm I}}\, {\dot \theta}_{0{\rm I} } (x) 
+ \oint Q_{{\rm I}}\, {\dot \phi}_{0{\rm I} } (x) 
- \int dt \, \H
\label{new9a}
\eeq
We have already obtained $d\mu$ and $\H$ for region I. With the fields in (\ref{new1}),
the constraints become
\beqar
\nabla \cdot E &=& \nabla^2 {\tilde \sigma}, \hskip .3in
\nabla \cdot A = \nabla^2 {\tilde \theta}\nonumber\\
\delta(\nabla\cdot E) &=& (\det(-\nabla^2)_{\rm I})^{-1}\, \delta ({\tilde\sigma})\nonumber\\
\delta(\nabla\cdot A)  &=& (\det(-\nabla^2)_{\rm I})^{-1}\, \delta ({\tilde\theta})
\label{new10}
\eeqar
This is equivalent to imposing Gauss law with test functions going to zero on the boundary $\del I$.
Using (\ref{new7}) and (\ref{new8}), the partition function (\ref{coulz}) can be now obtained as
 \beqar
 Z_{\rm I}&=&  \int d\mu_{\rm I} ~ \delta ({\tilde \sigma}_{\rm I})~\delta({\tilde \theta}_{\rm I})~(\det (-\nabla^2) )^{-1}_{\rm I} ~
 e^{i \S_{\rm I}}\nonumber\\
 &=& \int  [d \E d\theta_0]_{\rm I} \, [ d{\tilde \Pi} d B]_{\rm I} \, \exp\left({i \S_{\rm I}\vert_{{\tilde \sigma}, {\tilde \theta} = 0} }\right)
 \label{new11}
 \eeqar
 We can carry out the integration over $B$ as well to rewrite this as
 \beqar
 Z_{\rm I} &=& \int [d\E d\theta_0]_{\rm I} \, [d {\tilde \Pi}]_{\rm I} ~e^{i {\tilde S}_{\rm I}} 
 \nonumber\\
 {\tilde \S}_{\rm I} &=&\int {1\over 2} \left[ {\dot {\tilde\Pi}}_{\rm I}^2 - ( \nabla {\tilde \Pi}_{\rm I} )^2 \right]
 + \oint \left[ \E_{ {\rm I}}\, {\dot \theta_{0{\rm I} }} - {1\over 2} \E_{ {\rm I}} \, M^{-1}_{\rm I} \, \E_{{\rm I}}
 \right]
 \label{new12}
 \eeqar
 We can do a similar calculation, resulting in similar formulae, for region II. 
 

Now we want to consider the theory defined on the full space
I$\cup$II and consider integrating over
the degrees of freedom in region II. The resulting theory is to be compared to
the theory intrinsically defined in region I, namely to (\ref{new12}).
For the theory on the full space, we use the same parametrization of fields as
in (\ref{1}). Further we assume that the fields go to zero at the spatial boundary of the full space.
 This leads to
\beqar
\S &=& \int \left[ \del_i \sigma \, \del_i {\dot \theta} + \del_i \Pi \, \del_i {\dot \phi}
- {1\over 2} \left( (\del_i \sigma)^2 + (\del_i \Pi )^2 + B^2 \right) \right]\label{new12a}\\
Z_{\rm full} &=&\int [d\sigma d\theta] \, [d\Pi dB] \, [\det (-\nabla^2)]^2\,
\delta (\nabla\cdot E)\, \delta(\nabla\cdot A)~ e^{i \S}\nonumber\\
&=&\int [d\Pi ]~ \exp\left( { i \over 2}\int {\dot \Pi}^2 - (\nabla \Pi )^2 \right)
\label{new12b}
\eeqar
On the full space, we have the theory for a scalar field $\Pi$. However, instead of just considering this theory, we want to rewrite the action (\ref{new12a}) with the field variables decomposed 
into the two regions I and II.
This can be done by writing
\beq
\theta (x) =
\begin{cases} 
{\tilde \theta}_{\rm I} (x) + \oint_{\del {\rm I}} \theta_{0} (y)\,
\,n \cdot \del G (y,x)_{\rm I} &~~~~{\rm in ~I}\\
{\tilde \theta}_{\rm II} (x) + \oint_{\del {\rm II}} \theta_{0} (y)\,
\,n\cdot \del G (y,x)_{\rm II} &~~~~{\rm in~ II}\\
\end{cases}
\label{new16}
\eeq
Here $\theta_0$ is the value of $\theta$ on the interface between the two regions.
There are similar expressions for the other fields as well.
The action in terms of variables split into the two regions becomes
\beqar
\S_{\rm split} &=& \int_{\rm I} \left[ (-\nabla^2{\tilde \sigma}_{\rm I} ) \,{\dot{\tilde \theta}}_{\rm I} -{1 \over 2} (\del_i \tilde{\sigma}_{\rm I})^2 \right]+ \int_{\rm II} \left[
(-\nabla^2{\tilde \sigma}_{\rm II} )\, {\dot{\tilde \theta}}_{\rm II} -{1 \over 2} (\del_i \tilde{\sigma}_{\rm II})^2 \right] \nonumber\\
&&+ \int   \Pi {\dot B} - {1 \over 2} \left[ (\del_i \Pi )^2 + B^2 \right]~
+ ~\oint \sigma_0 (M_{\rm I} + M_{\rm II} ) \,{\dot \theta}_0   -
{1\over 2} \sigma_0 (M_{\rm I} + M_{\rm II} ) \,\sigma_0 \nonumber\\
&=&\int_{\rm I} \left[ (-\nabla^2{\tilde \sigma}_{\rm I} ) \,{\dot{\tilde \theta}}_{\rm I} -{1 \over 2} (\del_i \tilde{\sigma}_{\rm I})^2 \right]+ \int_{\rm II} \left[
(-\nabla^2{\tilde \sigma}_{\rm II} )\, {\dot{\tilde \theta}}_{\rm II} -{1 \over 2} (\del_i \tilde{\sigma}_{\rm II})^2 \right] \nonumber\\
&&+ \int   \Pi {\dot B} - {1 \over 2} \left[ (\del_i \Pi )^2 + B^2 \right]~
+ ~\oint \E \,{\dot \theta}_0   -
{1\over 2} \E (M_{\rm I} + M_{\rm II} )^{-1} \,\E 
\label{new16a}
\eeqar
where $\E = (M_{\rm I} + M_{\rm II} ) \, \sigma_0$. 
We have dropped the cross terms $\int\epsilon_{ij} \del_j \Pi \del_i {\dot \theta}$ and $\int\epsilon_{ij} \del_i \sigma \del_j {\dot \phi}$
since by continuity of the tangential derivative of $\Pi$ and $\sigma$ across the interface
the surface contributions cancel out. (Their inclusion will not change
anything that follows, except for the definition of $\E$ in terms of $\sigma_0$.
This is immaterial, we can just consider the redefined
$\E$ as the conjugate variable to $\theta_0$.)
The action for $\Pi, \, B$ will lead to the usual scalar field results, and since
our focus is on the factoring out of the gauge degrees of freedom, we
do not display the action for the $\Pi$ and $B$ fields in terms of variables
in each region. We will see how it reduces to a scalar field result.

The phase space volume element in these variables is
\beq
d\mu_{\rm split}  = [d{\tilde \sigma} d{\tilde \theta}]_{\rm I} \,[d{\tilde \sigma} d{\tilde \theta}]_{\rm II} \,\det (-\nabla^2)_{\rm I} \, \det (-\nabla^2)_{\rm II}\,
[d \E d\theta_0] \times d\mu_{\Pi, B}
\label{new16b}
\eeq
As for the expressions for the constraints in terms of these variables, 
the nature of the test functions
is the crucial ingredient. Considering test functions $f, ~h$, whose boundary values on the interface are $f_0, ~h_0$ respectively, and with the tilde-functions vanishing on the interface,
we have
\beqar
\int \del_i  f\, E_i &=& \int_{\rm I} \tilde{f }_{\rm I}(-\nabla^2 {\tilde \sigma}_{\rm I}) + 
 \int_{\rm II} \tilde{f}_{\rm II} (-\nabla^2 {\tilde \sigma}_{\rm II})  + \oint
 f_0 \, \E \approx 0\nonumber\\
\int \del_i  h\, A_i &=& \int_{\rm I} \tilde{h}_{\rm I} (-\nabla^2 {\tilde \theta}_{\rm I}) + 
 \int_{\rm II} \tilde{h}_{\rm II} (-\nabla^2 {\tilde \theta}_{\rm II})  + \oint
  h_0 \, (M_{\rm I} + M_{\rm II} )\,\theta_0  \approx 0
\label{new16c}
\eeqar
For the theory on the full space, $\theta$-dependence is eliminated everywhere
including the interface, so, based on (\ref{new16c}), we must interpret the constraints
as
\beq
\delta (\nabla \cdot E)~\delta(\nabla \cdot A) = \delta [ -\nabla^2{\tilde \sigma}_{\rm I}]\,\delta [ -\nabla^2{\tilde \sigma}_{\rm II}]\,
\delta [ \E ]~ \delta [ -\nabla^2{\tilde \theta}_{\rm I}]\,\delta [ -\nabla^2{\tilde \theta}_{\rm II}]\,
\delta [(M_{\rm I} + M_{\rm II})  \theta_0 ]
\eeq
Further, we can use the splitting formula
(or the BFK gluing formula \cite{BFK})
\beq
\det (-\nabla^2) = \det (-\nabla^2)_{\rm I} \, \det (-\nabla^2)_{\rm II} \, \det (M_{\rm I}
+ M_{\rm II})
\label{new17}
\eeq
Using (\ref{new16b})-(\ref{new17}), it is then easy to verify that
\beqar
&&\int d\mu_{\rm split} \,
\delta [ -\nabla^2{\tilde \sigma}_{\rm I}]\,\delta [ -\nabla^2{\tilde \sigma}_{\rm II}]
\delta [ \E ]\, \delta [ -\nabla^2{\tilde \theta}_{\rm I}]\,\delta [ -\nabla^2{\tilde \theta}_{\rm II}]
\delta [(M_{\rm I} + M_{\rm II})  \theta_0 ] \, \det (-\nabla^2) ~e^{i \S_{\rm split}}\nonumber\\
&&\hskip .2in= \int [d\Pi ]~ \exp\left( { i \over 2}\int {\dot \Pi}^2 - (\nabla \Pi )^2 \right)
\label{new17a}
\eeqar
does indeed reproduce the partition function in (\ref{new12b}). The calculations from
that point until (\ref{new17a}) were meant to show that the parametrization
with the splitting as in (\ref{new16}) does capture the theory on the full space.

Consider now the integration of the degrees of freedom in region ${\rm II}$.
From the point of view of the theory in ${\rm II}$, the modes due to
$\E$, $\theta_0$ are physical edge degrees of freedom, they are not considered 
as gauge degrees of freedom. This means that one integrates
over only the ${\tilde \sigma}_{\rm II}$ and ${\tilde \theta}_{\rm II}$
without imposing the constraints which eliminate the edge degrees of freedom.
The corresponding test functions $f$, $h$ in (\ref{new16c}) are taken to vanish on the
interface, so that we get
\beqar
&&\int d\mu_{\rm split} \,
\delta [ -\nabla^2{\tilde \sigma}_{\rm I}]\,\delta [ -\nabla^2{\tilde \sigma}_{\rm II}]
 \delta [ -\nabla^2{\tilde \theta}_{\rm I}]\,\delta [ -\nabla^2{\tilde \theta}_{\rm II}]
\, \det (-\nabla^2) ~e^{i \S_{\rm split}}\nonumber\\
&&\hskip .3in= \det (M_{\rm I} + M_{\rm II}) \int  [d{\tilde \sigma} d{\tilde \theta}]_{\rm I} 
\, \delta [ -\nabla^2{\tilde \sigma}_{\rm I}]\,  \delta [ -\nabla^2{\tilde \theta}_{\rm I}]\,
\left[ \det (-\nabla^2)_{\rm I}\right]^2\,
[d\E d\theta_0] d\mu_{\Pi, B}\, e^{i \S}\nonumber\\
&&\hskip .3in= \det (M_{\rm I} + M_{\rm II}) \int  
[d\E d\theta_0] d\mu_{\Pi, B}\, e^{i \S}
\label{new17b}\\
&&\hskip .12in \S = \int  \left[ \E\, {\dot \theta_0} - {1\over 2}  \E ( M_{\rm I}  + M_{\rm II} )^{-1}\,
\E  \right] + \S_{\Pi, B}
\label{new17c}
\eeqar
This has exactly the structure we expect for the partition function in region ${\rm I}$, namely,
(\ref{new12}), except for the prefactor of $\det (M_{\rm I} + M_{\rm II})$.
Even though the integrations in (\ref{new12}) involve
$\E_{\rm I}$ and $\theta_{0{\rm I}}$ while we have
$\E$, $\theta_0$ in (\ref{new17b}), the result is identical once the integral is performed;
the result does not depend on the interface. 
In fact, defining a new variable $\xi =  (K^{-1})^{\half}  \theta_0$, where $ K = (M_{\rm I}
+ M_{\rm II})^{-1}$ or $M_{\rm I}^{-1}$ appropriately, we see that
\beq
\int [d\E d\theta_0] \exp\left( i \int \left[ \E \, {\dot \theta_0} - {\half} \E\, K\, \E \right]
\right) = {\rm constant} \int [d\xi] \exp\left( - {i\over 2} \int {\dot \xi}^2 \right)
\label{new20}
\eeq
where the constant does not depend on $K$. 
Also, in (\ref{new17b}, {\ref{new17c}) we have not displayed the integration over
 the $\Pi$-$B$-fields for
region II. Since the action for this part is that of a scalar field (which is $\Pi$),
we take this to be done as in the case of a scalar field.

So the only extra factor in reducing the theory by integrating over region ${\rm II}$, but
keeping the edge modes for ${\rm II}$, is $ \det (M_{\rm I} + M_{\rm II} )$.
This term arises from the phase volume and hence must be counted as
a degeneracy factor due to the additional modes.
The corresponding density matrix must be defined to account for the extra degeneracy as 
 \beq
 \rho = {{\mathbb 1}  \over \det (M_{\rm I} + M_{\rm II})}\, ~(\rho_{\Pi})_{\rm red}
 \label{glue5}
 \eeq
 where $(\rho_\Pi)_{\rm red}$ is the normalized reduced
 density matrix for a massless scalar (from the $\Pi$-$B$ sector)
 and ${\mathbb 1} $ is a unit matrix such that $\Tr\, {\mathbb 1}  = \det (M_{\rm I} + M_{\rm II})$. This degeneracy factor affects EE which will now be given by
 \beq
 S_E = S_{E\,\Pi} + \log \det (M_{\rm I} + M_{\rm II}) = \frac{{\mathbb A}}{12\sqrt{\pi}}\left(\frac{1}{\e} + \mathcal{O}(\e)\right) +  \Tr \log (M_{\rm I} + M_{\rm II})
 \label{glue6}
 \eeq
 where $\epsilon$ is the UV cutoff and ${\mathbb A}$ is the ``area'' of the entangling surface. (Once again, in 2+1 dimensions, this is just the length of the entangling surface.) 
 The first term $S_{E\,\Pi}$ is the contribution from the scalar field
 $\Pi$.
 The second term on the right hand side is the contact term.\footnote{There is a slight abuse of notation between (\ref{new17b}) and (\ref{glue5}) or 
  (\ref{glue6}). The determinant in (\ref{new17b}) involves a product over all time,
  which is not shown explicitly.
This is because we have a factor $\det (M_{\rm I} + M_{\rm II})$ on each time-slice, but,
since the operator does not involve time-derivatives, this gives an overall integration
over time in $\Tr \log (M_{\rm I} + M_{\rm II})$.
The factors $\det (M_{\rm I} + M_{\rm II})$  in (\ref{glue5}) and (\ref{glue6}) are at fixed time. }
 If we take
the regions I and II to be the left and right half-planes, then $M_{\rm I} \sim \sqrt{q^2}$,
$M_{\rm II} \sim \sqrt{q^2}$, where $q$ is the momentum along the interface \cite{cas1}.
In this case, we find
\beqar
\Tr \log  (M_{\rm I} + M_{\rm II}) &=& {1\over 2} \Tr \log q^2 + \Tr \log 2\nonumber\\
&=& -\frac{\mathbb A}{4\sqrt{\pi}}\int_{\epsilon^2}^\infty ds \frac{e^{-m^2s}}{s^{3/2}}
= - {{\mathbb A} \over 2 \sqrt{\pi}} \left( {1\over \epsilon} + {\cal O} (\epsilon ) \right)
\label{glue7}
\eeqar
Here we absorbed the $\Tr \log 2$ part into a redefinition of the cut-off and used
a mass term (i.e. $q^2 \rightarrow q^2 + m^2$) as an infrared regulator, although 
this is ultimately not needed for the answer displayed.
The result (\ref{glue7}) agrees with Kabat's calculation of the contact term. 
Notice that ${\half} \Tr \log q^2$ is the negative of the free energy of a massless
scalar in $d-2$ dimensions confined to the entangling surface, where $d$ is the
spacetime dimension of the theory.

To recapitulate, we see that the ``extractable part"  of vacuum 
entanglement is captured by a single massless scalar. 
If one eliminates the gauge degrees of freedom over the full space
${\rm I} \cup {\rm II}$, and then considers integrating over the $\Pi$, $B$ degrees of freedom in
region II, then this is all there is to the entropy.
However, if one keeps the $\E$, $\theta_0$ modes on the interface, since they are
physical from the point of view of the theory in region II,
then there is an additional contribution
 from the degeneracy.  This reproduces the contact term 
 obtained in the replica method. 
 Although we phrased the arguments in terms of the phase space functional integral,
 the key point is the splitting of the phase volume. The factor $\det (-\nabla^2)$, which 
 may be viewed as arising from factoring out the gauge degrees of freedom,
 does not trivially factorize into the two regions. The ``extra piece"
 $\det (M_{\rm I} + M_{\rm II})$ is precisely the same surface term needed for the BFK 
 gluing formula (\ref{new17}). We identify this as
 the contact term.
Also, even though we have considered a flat spacetime with a 
flat interface between the two regions,
it is clear that the result can be generalized to
any bipartite partition of space, with the appropriate $M_{\rm I}$ and $M_{\rm II}$.

We will close this section with a few comments.
In going from (\ref{new11}) to (\ref{new12}), we integrated over $B$. 
The resulting path integral is thus appropriate for describing the evolution of
$E$-diagonal wave functions, since $\Pi$ is part of the electric field.
One could also consider integrating over $\Pi$ to obtain a
$\phi$-diagonal representation with the corresponding wave functions as functions of
$\phi$. This results in a determinant $(\det (-\nabla^2))^{-{\half}}$ with the relevant
part of the action as
\beq
\S = {1\over 2} \int \left[{\dot \phi}\, (-\nabla^2) \, {\dot \phi}  - (-\nabla^2 \phi )^2\right]
\label{new21}
\eeq
(Here we consider the full space for simplicity.)
Naively, it would seem that this does not lead to a scalar field result for the EE, since there are
higher derivatives involved. However, notice that
the commutation rules are
\beq
 [ \phi (x), \, {\dot \phi} (y) ] = i \, G(y,x)
 \label{new22}
 \eeq 
If we consider splitting the manifold into two regions, say, I and II, with the 
 corresponding $\phi_{\rm I} $ and $\phi_{\rm II}$, then this
 commutation rule tells us that there is some entanglement since
 $[ \phi_{\rm I}, {\dot \phi}_{\rm II} ] \neq 0$ due to the nonlocality of the Green's function; there is an uncertainty principle for simultaneous measurements of $\phi$ and ${\dot \phi}$ for far separated regions. This is true irrespective of which state of the system (or wave function) we choose 
and could be an additional source of entanglement beyond what is obtained from the 
wave function. Calculations just using the wave function are not adequate.
To simplify the analysis, one option is to choose variables which give local
commutation rules, thereby transferring all entanglement to the wave function.
One such choice is
\beq
\vf = \sqrt{-\nabla^2}\, \phi
\label{new23}
\eeq
In this case, the action for the $\phi$-part reduces to
\beq
\S = {1\over 2} \int \left[ ({\dot \vf})^2  - (\nabla \vf )^2\right]
\label{new24}
\eeq
Since $[d\phi] (\det (-\nabla^2))^{-{\half}} = [d\vf]$, the measure of integration also correctly corresponds
to what is needed for a scalar field. Thus the previous results are still obtained.

It is possible to include edge modes on the boundary of
the full space as well, although
they are not important for the entanglement entropy. 
The Hamiltonian for the full space has a form similar to (\ref{new8})
 (without the subscript I, of course).
The term involving
$\E$ is the term corresponding to the edge modes. 
Since the $\E$ at different points on the boundary commute
at equal time, the $\E$-dependent term in the Hamiltonian
is like a free particle kinetic energy term and gives
continuous eigenvalues.
This is in agreement with \cite{bal}.
 
\section{ The Maxwell-Chern-Simons theory}

We shall now consider a similar analysis for the Maxwell-Chern-Simons (MCS) theory.
The action is given by
\beq
\S_{\rm MCS} = \int d^3x~ \left[{1\over 2}( E^2 - B^2) + {k e^2 \over 4 \pi} \epsilon^{\mu\nu\alpha}
A_\mu \del_\nu A_\alpha \right]
\label{15}
\eeq
where $e$ is the coupling constant.
While the Maxwell term is manifestly gauge invariant, the Chern-Simons (CS) term
changes by a boundary term upon carrying out a gauge transformation.
This boundary term will have
a contribution involving the spatial boundary and two terms on the
initial and final time-slices which results from the time-integration.
The latter terms
will be part of the Gauss law of the theory, while the spatial boundary contributions will
vanish for those transformations which become the identity at the spatial boundary.
We will consider only such transformations for the boundary of the full
space,
so that the CS term can be taken to be gauge invariant in the full space. 

As in the case of the Maxwell theory, we want to start with the
theory defined on the full space ${\rm I} \cup {\rm II}$ and write it on terms of
variables defined on each region.
Again, we choose the $A_0 = 0 $ condition.  We can simplify the canonical structure and the
Hamiltonian in the parametrization we use and then consider
integrating out the degrees of freedom in one of the two regions, say, II.
Using variables in the full space, but split up as in (\ref{new16}), we find
\beqar
\A&=&\int \left[ E_i \, \delta A_i - {m \over 2} \epsilon_{ij} A_i \, \delta A_j \right] \nonumber\\
&=&\int_{\rm I}  \bigl[(- \nabla^2 {\tilde \sigma}_{\rm I} ) + m \nabla^2 {\tilde \phi}_{\rm I} \bigr]\, \delta{\tilde \theta}_{\rm I} +
\int_{\rm II}  \bigl[(- \nabla^2 {\tilde \sigma}_{\rm II} ) + m \nabla^2 {\tilde \phi}_{\rm II} \bigr]\, \delta{\tilde \theta}_{\rm II} + \int \del_i \Pi \del_i \delta \phi\nonumber\\
&&+ \oint \sigma_0  (M_{\rm I} + M_{\rm II} ) \delta\theta_0 + + \delta \left[ {m \over 2}
\int \del_i \phi \, \del_i \theta \right]\nonumber\\
&=&\int_{\rm I}  (- \nabla^2 {\tilde \alpha}_{\rm I} ) \, \delta{\tilde \theta}_{\rm I} +
\int_{\rm II}  (- \nabla^2 {\tilde \alpha}_{\rm II} )  \, \delta{\tilde \theta}_{\rm II} +\int \del_i \Pi \del_i \delta \phi
+ \oint \alpha_0 (M_{\rm I} + M_{\rm II} ) \delta\theta_0 \nonumber\\
&&\hskip .1in + \delta \left[ {m \over 2} 
\int \del_i \phi \, \del_i \theta \right]
\label{new25}
\eeqar
where $\alpha = \sigma - m \phi$, $m = ke^2/2\pi$. 
In arriving at this expression, we have also
dropped some terms which cancel out between the two regions
due to the continuity of the fields,
\beq
\left[ \del_\tau \Pi_{0{\rm I}} + {m \over 2} \del_\tau \theta_{0{\rm I}}  \right]\, \delta \theta_{0{\rm I}} 
- \left[ \del_\tau \Pi_{0{\rm II}} + {m \over 2} \del_\tau \theta_{0{\rm II}}  \right]\, \delta \theta_{0{\rm II}} 
= 0
\label{new25a}
\eeq
The last term in (\ref{new25}) is a canonical transformation, so
it gives the well-known phase factor for the wave functions of the MCS theory.
It will not be important for our discussion of the entanglement entropy.
(In Appendix A we write down $\A$ and simplify it, showing one can choose
$\phi_0 = 0$. We have used the resulting expression along with
(\ref{new25a}) to obtain (\ref{new25}).)
The Hamiltonian can be simplified as
\beqar
\H_{\rm MCS} &=& {1\over 2} \int ( E^2 + B^2 )\nonumber\\
&=&{1\over 2} \int_{\rm I}  \bigl[ (- \nabla^2 {\tilde\sigma}_{\rm I}) \, {\tilde \sigma}_{\rm I}\bigr]
+ {1\over 2}  \int_{\rm II} \bigl[(- \nabla^2 {\tilde\sigma}_{\rm II}) \, {\tilde \sigma}_{\rm II} \bigr]
+ {1\over 2} \oint \sigma_0  (M_{\rm I} + M_{\rm II}) \, \sigma_0 ~+~ \H^{(0)}_{\Pi, B}
\label{new26}
\eeqar
The $\Pi$-$B$ part of the Hamiltonian corresponds to a scalar field and is given by
\beq
\H^{(0)}_{\Pi, B} = {1\over 2} \int  \left[ ( \nabla \Pi)^2 +  ({-\nabla^2 \phi})^2 \right] 
\label{new27}
\eeq
Denoting $M = M_{\rm I} + M_{\rm II}$,
the $\sigma$-dependent terms of the Hamiltonian can be written in terms of
$\alpha$ as
\beqar
\H_{\rm MCS} &=&  {m^2\over 2} \left[ \int_{\rm I}  \bigl[ (- \nabla^2 {\tilde\phi}_{\rm I}) \, {\tilde \phi}_{\rm I}\bigr]
+  \int_{\rm II} \bigl[(- \nabla^2 {\tilde\phi}_{\rm II}) \, {\tilde \phi}_{\rm II} \bigr]
 \right]\nonumber\\
&&
+{1\over 2} \int_{\rm I}  \bigl[ (- \nabla^2 {\tilde\alpha}_{\rm I}) \, {\tilde \alpha}_{\rm I}\bigr]
+ {1\over 2}  \int_{\rm II} \bigl[(- \nabla^2 {\tilde\alpha}_{\rm II}) \, {\tilde \alpha}_{\rm II} \bigr]
+ {1\over 2} \oint \alpha_0  M\, \alpha_0 \nonumber\\
&&+ m \left[\int_{\rm I} \nabla{\tilde \alpha} \nabla{\tilde \phi}  +  \int_{\rm II} \nabla{\tilde \alpha} \nabla{\tilde \phi}  \right]
\label{new28}
\eeqar
Notice that the first line of the right hand side involving only $\phi$-dependent terms
can be combined into the full space integral again.
(If we retained $\phi_0$, there would be an additional term
$\int \phi_0 M \phi_0$, which would be just what is needed to combine the terms
into the full space integral.)
Thus
\beqar
\H_{\rm MCS} &=&
{1\over 2} \int_{\rm I}  \bigl[ (- \nabla^2 {\tilde\alpha}_{\rm I}) \, {\tilde \alpha}_{\rm I}\bigr]
+ {1\over 2}  \int_{\rm II} \bigl[(- \nabla^2 {\tilde\alpha}_{\rm II}) \, {\tilde \alpha}_{\rm II} \bigr]
+ {1\over 2} \oint \alpha_0  M\, \alpha_0\nonumber\\
&&+ m \left[\int_{\rm I} \nabla{\tilde \alpha} \nabla{\tilde \phi}  +  \int_{\rm II} \nabla{\tilde \alpha} \nabla{\tilde \phi}  \right] ~+~ \H_{\Pi, B}\nonumber\\
\H_{\Pi, B} &=& {1\over 2} \int  \left[ ( \nabla \Pi)^2 +  ({-\nabla^2 \phi})^2  + m^2 (\nabla \phi )^2\right] 
\label{new29}
\eeqar
Notice that $\H_{\Pi, B}$ corresponds to a massive scalar field.
This part of the theory will contribute to the EE as a massive scalar field.

The volume element for the phase space can be obtained from (\ref{new25})
as
\beq
d\mu_{\rm split}  = [d{\tilde\alpha} d{\tilde \theta}]_{\rm I} \, [d{\tilde\alpha} d{\tilde \theta}]_{\rm II} 
\,[d \alpha_0 \, d\theta_0] \, \det (-\nabla^2)_{\rm I} \, \det (-\nabla^2)_{\rm II}\,
\det M
\label{new30}
\eeq
The constraint corresponding to the Gauss law can be 
identified from (\ref{new25}) and reads
\beq
 \int_{\rm I} \tilde{f }_{\rm I}(-\nabla^2 {\tilde \alpha}_{\rm I}) + 
 \int_{\rm II} \tilde{f}_{\rm II} (-\nabla^2 {\tilde \alpha}_{\rm II})  + \oint
 f_0 \, M \alpha_0 \approx 0\
\eeq
Thus, if
we want to factor out $\theta$
on the full space including the interface, the constraints
are given by
\beq
\C  = \delta [ - \nabla^2 {\tilde \alpha}_{\rm I}] \, 
\delta [ - \nabla^2 {\tilde \alpha}_{\rm II}] \, \delta[ M \alpha_0]
\times  \delta [ - \nabla^2 {\tilde \theta}_{\rm I}] \, 
\delta [ - \nabla^2 {\tilde \theta}_{\rm II}] \, \delta[ M \theta_0]
\label{new31}
\eeq
The first set of terms on the right hand side correspond to the Gauss law
while the second set gives the Coulomb gauge-fixing conditions.
The Poisson bracket of the two constraints is again $- \nabla^2$
which may be split up using the BFK formula (\ref{new17}).
It is then easy to see that the theory on the full space reduces to
the $\Pi$-$B$ sector, i.e., the theory of a massive scalar field.

However, as discussed before,
in integrating over the region II, the modes $\alpha_0$ and $\theta_0$ become physical degrees of freedom. We integrate only over $\tilde{\alpha}_{\rm II}$ and $\tilde{\theta}_{\rm II}$ without imposing the constraints which eliminate the edge degrees of freedom. Following the same procedure as in the Maxwell case, we find that we get an extra factor of $\det M$ which can be essentially interpreted as the contact term. 
Thus the EE contributed by the gauge fields is 
\beq
S_E = S_E(m) +  \Tr \log (M_{\rm I} +
M_{\rm II})
\label{new31a}
\eeq
$S_E(m)$ is identical to the EE of a massive scalar field with mass $ m = \frac{ke^2}{2\pi}$. When the entangling surface is flat, i.e., a planar or straight line interface,
 one can explicitly evaluate this term to get
\beq
S_E(m ) = {{\mathbb A}\over 12 \sqrt{\pi}} \left( {1\over \epsilon}
- {k e^2 \over 2 \pi}  \sqrt{\pi} + \O (\epsilon )\right) 
\label{new31b}
\eeq
which brings out the massive corrections to the pure Maxwell case studied earlier. Most notably, we see the presence of a cut-off independent finite term proportional to the mass-gap that scales as the area of the entangling surface.
$ \Tr \log (M_{\rm I} +
M_{\rm II})$ is again the contact term, which is formally identical to what is obtained for the (massless) Maxwell case.

When the boundary between region I and II has nontrivial topology, such as a circle, there can be an additional contribution to the EE, depending on the procedure of integrating out fields
in II. To see how this can arise, we first consider splitting the fields as in (\ref{new16}), but keep distinct
values $\theta_{0 {\rm I}}$, $\theta_{0 {\rm II}}$ on the two sides of the interface,
\beq
\theta (x) =
\begin{cases} 
{\tilde \theta}_{\rm I} (x) + \oint_{\del {\rm I}} \theta_{0{\rm I}} (y)\,
\,n \cdot \del G (y,x)_{\rm I} &~~~~{\rm in ~I}\\
{\tilde \theta}_{\rm II} (x) + \oint_{\del {\rm II}} \theta_{0{\rm II}} (y)\,
\,n\cdot \del G (y,x)_{\rm II} &~~~~{\rm in~ II}\\
\end{cases}
\label{new31c}
\eeq
with a similar result for the other fields.
The terms in the canonical one-form
relevant to the fields $\alpha_0$, $\theta_0$ are
\beqar
\A (\alpha_0, \theta_0 ) &=&
\oint \left[\E_{\rm I}\, \delta \theta_{0{\rm I}} + {m \over 2}\, \del_\tau \theta_{0{\rm I}}
\delta \theta_{0{\rm I}} \right]+
\oint \left[ \E_{\rm II} \, \delta \theta_{0{\rm II}} - {m \over 2} \,\del_\tau \theta_{0{\rm II}}
\delta \theta_{0{\rm II}} \right]\nonumber\\
\E_{{\rm I}/{\rm II}} &=&  \alpha_{0{{\rm I}/{\rm II}} }\, M_{{\rm I} /{\rm II}} \pm \del_\tau \Pi_{0{{\rm I}/{\rm II}}}
\label{new31d}
\eeqar
We can consider the symplectic reduction of this via the constraints
$\theta_{0{\rm I}} - \theta_{0{\rm II}} \approx 0$,
$\E_{{\rm I}} - \E_{{\rm II}} \approx 0$, which are the matching conditions at the interface.
Clearly $\A$ reduces to the previous expression (\ref{new25}) in this case. 
The $\del_\tau \Pi_0$ and
$\del_\tau \theta_0$ terms all cancel out between the two regions.
The phase volume also
reduces correctly. From (\ref{new31d}) we get
\beq
d\mu = [d\E_{{\rm I}} d\theta_{0{\rm I}}] \,[d\E_{{\rm II}} d\theta_{0{\rm II}}] \,
\det M_{\rm I} \, \det M_{\rm II} 
\label{new31e}
\eeq
The Poisson bracket of the constraints is
\beq
\{ \theta_{0{\rm I}} - \theta_{0{\rm II}} , \E_{{\rm I}} - \E_{{\rm II}}
\} = M^{-1}_{\rm I} + M^{-1}_{\rm II}
\label{new31f}
\eeq
so that the reduced volume is
\beq
\int_{\E_{{\rm II}}, \theta_{0{\rm II}}} d\mu \, \delta [ \E_{{\rm I}} - \E_{{\rm II}}]
\, \delta[ \theta_{0{\rm I}} - \theta_{0{\rm II}} ] \det \left( M^{-1}_{\rm I} + M^{-1}_{\rm II} \right)
= [d\E d\theta_0 ] \det (M_{\rm I} + M_{\rm II} )
\label{new31g}
\eeq
Thus for a planar interface, we do recover the previous result, with the contact term as
$\Tr \log (M_{\rm I} + M_{\rm II})$ \footnote{Since we have $d\Pi d\phi$ as well in the measure,
$d\E \wedge d\Pi d\phi = d\alpha_0\wedge d\Pi d\phi$, so we can drop
the $\del_\tau\Pi$ part at this point.}.

In the case of an interface which is a circle (or has the topology of a circle), the condition
$\theta_{0{\rm I}} - \theta_{0 {\rm II}} \approx 0$ is too restrictive.
Since $\theta_0$ is an angular variable, it can shift by
an integer multiple of $2 \pi$ upon going around the circle, so that
we only need
\beq
\theta_{0{\rm I}} - \theta_{0 {\rm II}} \approx 0 \mod 2 \pi {\mathbb Z}
\label{new31h}
\eeq
One way to enforce this constraint is to add a term $\H_{constraint}$ to the Hamiltonian,
\beq
\H_{constraint} = {\lambda \over 2 \pi}  \oint \left[ 1 - \cos (\theta_{0{\rm I}} - \theta_{0{\rm II}})
\right]
\label{new31i}
\eeq
with $\lambda \rightarrow \infty$ eventually. The result for the EE will thus be
of the form
\beq
S_{\rm MCS}  = S_E(m) +  \Tr \log (M_{\rm I} +
M_{\rm II})+ S_{Chiral}(k) \label{smcs}
\eeq
where $S_{Chiral}(k)$ refers to the contribution from integrating over 
$\E_{{\rm I}}$, $\E_{{\rm II}}$, $\theta_{0{\rm I}}$,
$\theta_{0{\rm II}}$ with the constraint $\delta[ \E_{{\rm I}} - \E_{{\rm II}}]$
and the term (\ref{new31i}).
For the pure Chern-Simons case (without the Maxwell action),
such a calculation has been done
\cite{CSE1,tope,CFTE, PIE}.
The key result of that calculation is that there is a topological contribution
$- {\half} \log k$ in $S_{Chiral}(k)$ in addition to the usual regularization-dependent
terms.
The topological term arises purely from a set of ``zero modes" in
the expansion of $\theta_0$ and we can show that the same result
holds for the 
Maxwell-Chern-Simons theory as well. In other words,
\beq
S_{Chiral, MCS} (k) = - {1\over 2} \log k + \cdots
\label{new31j}
\eeq
where the ellipsis again refers to regularization-dependent terms.
This result is shown in Appendix B.

\section{Yang-Mills Theory}

We will now turn to the issues in computing the EE for the nonabelian gauge theory in 2+1 dimensions. It is useful to start with considerations
within a perturbative scheme.
In this case, we can 
consider the phase space functional integral which is given in the Coulomb gauge by
\beq
Z = \int [dE \, dA] \, \delta (D_i E_i ) \, \delta (\del\cdot A) \, \det (- \del\cdot D) \, e^{i \S}
\label{nonab1}
\eeq
Similar to the situation with the Abelian theory, we can introduce the parametrization
\beq
A_i^a = \del_i \theta^a + \epsilon_{ij} \del_j \phi^a , \hskip .3in
E_i^a =  \del_i {\sigma}^a + \epsilon_{ij} \del_j \Pi^a
\label{nonab2}
\eeq
This is not the parametrization best-suited to the nonabelian theory, nevertheless
we can, in principle, consider this as a starting point for perturbation theory.
A mode decomposition for fields in regions I and II can be done in a way analogous to
(\ref{new1}) and (\ref{new2a}).
The action will contain terms which
 mix the boundary fields and the bulk fields, and the integration over various fields will have
 to be done in a perturbative expansion.
The determinants involved in the factorization of the phase volume, i.e., the
extension of formula (\ref{new17}),
will also have to be
obtained via a similar expansion.
To the lowest order, the results will coincide with the Maxwell theory except 
for a multiplicative factor of dim$G$ (= $N^2 -1$ for $SU(N)$), since we have
dim\,$G$ fields rather than one.

This approach is clearly unsatisfactory since we do not see any nonperturbative effects
in the entropy. Some nonpertrubative effects, such as the mass gap, can be included
using the KKN approach to gluodynamics \cite{KKN}. Again, we do not expect an exact
calculation, but there is a qualified {\it free} limit of the theory which corresponds to 
the inclusion of the nonperturbative mass gap but otherwise ignores the interactions. So what we need is a formulation where we can use the mass term and expand around this 
{\it free} limit to get further corrections.
We shall refer to \cite{KKN} for the relevant technical details, capturing only what is relevant for the computation of EE below.

As usual, we start with the choice of $A_0 = 0$.
The nonabelian analog of the fields $\theta$, $\phi$, for $SU(N)$ gauge symmetry,
are $SL(N, \mathbb{C})$-valued complex matrices $M$ and $M^\dagger$ which parametrize the gauge fields as 
\beq
A = {\half} (A_1 + i A_2) =
-\partial MM^{-1}, \hspace{.3cm} \bar A = {\half} (A_1 -i A_2) = M^{\dagger -1}\bar \partial M^{\dagger}\label{M}
\eeq
Under gauge transformations, $M \rightarrow g \,M$. The hermitian matrix $H = M^\dagger M$, 
which is in $ SL(N, \mathbb{C})/ SU(N)$, 
provides a coordinatization of the space of gauge-invariant configurations $\mathcal{C}$ and can be regarded as the basic gauge-invariant observable (the nonabelian analog of $\phi$). The measure on the configuration space is given by \cite{KKN}
\beq
d\mu_{\mathcal C}= d\mu[H] ~e^{2\, c_A\, \S_{WZW}[H]}\label{measureH}
\eeq
where $d\mu [H]$ is the Haar measure on the space of hermitian matrices
$H$ and
\beq
\S_{WZW}[H] = \frac{1}{2\pi}\int \Tr(\partial H \bar \partial H^{-1}) +\frac{i}{12\pi}\int \epsilon^{\alpha \beta \gamma} \Tr(H^{-1}\partial_\alpha HH^{-1}\partial_\beta HH^{-1}\partial_\gamma H) 
\label{nonab3}
\eeq
is the Wess-Zumino-Witten (WZW) action for the field $H$. Also
$c_A$ in (\ref{measureH}) is the adjoint Casimir of the group, equal to $N$ for $SU(N)$. As shown in \cite{KKN}, the $S_{WZW}$ factor
arises from the Jacobian for change of variables from $A,\bar A$ to $H$. The Yang-Mills Hamiltonian can be rewritten in terms of the gauge invariant variable $H$, or more conveniently in terms of the currrent $J = \frac{c_A}{\pi}\partial H H^{-1}$, as
\beq
{\mathcal H} = \frac{e^2c_A}{2\pi}\left(\int J^a\frac{\delta}{\delta J^a} + \int \Omega ^{ab}(x, y)\frac{\delta}{\delta J^a(x)}\frac{\delta}{\delta J^b(y)}\right)  + \frac{2\pi^2}{e^2c_A^2}\int(\bar \partial J^a \bar \partial J^a) 
\label{HJ}
\eeq
where
\beq
\Omega^{ab}(x,y) = [\mathcal{D}_x\bar \G(y,x)]^{ab}, \hspace{.5cm} \mathcal{D}_x = \frac{c_A}{\pi}\partial_x\delta^{ab} + if^{abc}J^c(x)
\label{nonab4}
\eeq
The Hamiltonian involves the covariant Green's function $\bar \G$ and hence needs to be defined with appropriate regulators in place. We refer to the original papers \cite{KKN} for these and other technical issues. For our purposes, it is important to highlight that the first term in the Hamiltonian acts as a mass term when acting on functionals of $J$.  It is this term that renders the theory massive and its coefficient $
m = ({e^2c_A}/{2\pi})$
is the basic mass gap of the nonabelian theory. Furthermore, (\ref{HJ}) is self-adjoint only with respect to the measure (\ref{measureH})  and the coefficient of the WZW action is fixed by the requirement of self-adjointness.

While  (\ref{HJ})  is not known to be exactly solvable, one can compute the ground state wave functional  in a strong coupling expansion. This has been carried out in a series of papers, and the resulting string tension compares remarkably well with lattice results \cite{{KKN4}, {string-tension}}. To the
leading order in this expansion, the wave functional is
\beq
\Psi = \exp \left(-\frac{2\pi^2}{e^2c_A^2}\int \bar\partial J \frac{1}{m+\sqrt{m^2 - \nabla ^2}}\bar\partial J + \mathcal{O}(J^3)\right)
\label{gaussl}
\eeq  
Focusing on just the quadratic part of the wave functional, we can rewrite it in more familiar terms. One can parametrize $H$ as $H = e^\Phi$ and absorb the exponential factor
involving $S_{WZW}[H]$ in (\ref{measureH}) in a redefinition of the wave function. After expanding to quadratic order in $\Phi^a$ and redefining $\Phi^a = \frac{1}{\sqrt{-\nabla^2}}\vf^a$, we get
\beq
\Psi = \exp \left( -\frac{1}{2}\int \vf^a \sqrt{m^2 - \nabla ^2} \vf^a + \cdots\right) 
\label{gaussm}
\eeq
The wave function (\ref{gaussl}) is square integrable with the integration measure
for $H$ given by (\ref{measureH}), while the wave function
(\ref{gaussm}) is square-integrable with just the Haar measure $d\mu [H]$.
The same manipulations allow us to rewrite the Hamiltonian in terms of its action on
(\ref{gaussm}) as
\beq
{\mathcal H} = -\frac{1}{2}\int \frac{\delta ^2}{\delta \vf^a \delta \vf^a} + \frac{1}{2}\int \vf^a (-\nabla ^2 + m^2) \vf^a + \cdots 
\label{gauss-ham} 
\eeq
The ellipsis refer to  cubic and higher order terms in $\vf$. Ignoring the higher order terms, it is clear that (\ref{gaussm}) is the wave functional for (\ref{gauss-ham}) . This quadratic theory is the non-standard {\it free} limit of the nonabelian theory we alluded to earlier. In this approximation, the gauge theory decouples into dim\,$G$ copies of a massive scalar theory. 

Having reduced the problem to that of dim$G$ copies of a massive scalar, we can borrow from
(\ref{sm}) and write the
EE for this theory (in the nonstandard {\it free} limit mentioned above) as
\beq
S_E = {\rm dim}G\, \frac{{\mathbb A}}{12\sqrt{\pi}}\left(\frac{1}{\e} - {e^2 c_A \over 2 \pi}\,
\sqrt{\pi} + \mathcal{O}(\e)\right)
 \label{sm2}
\eeq

A couple of comments are in order at this point.
\begin{enumerate}
\item The first term in (\ref{sm2}) corresponds to the EE for dim\,$G$ copies of the Maxwell theory,
 see the first term of the expression for $S_E$ in (\ref{glue6}).
\item Apart from the divergent $1/\e$ term, we see that the three-dimensional entropy also contains a finite negative term $- {\rm dim}G\,({\mathbb A}\,m /{12})$ that is proportional to the mass gap. This is reminiscent of the topological contribution to the entropy in Chern-Simons theories but, unlike the topological term, this contribution scales as the area. This result shows a direct link between the finite part of EE and the volume measure on the gauge theory configurations which - in turn - is deeply connected to IR properties of the theory.  
Specifically, $m$ is renormalized in the presence of  an explicit or induced Chern-Simons term by a finite amount \cite{renk},
\[
m \rightarrow m + \frac{e^2k}{4\pi}
\]
where $k$ is the Chern-Simons level number.
In theories with extended supersymmetry, the induced level number
exactly cancels $m$ and the mass gap is renormalized to zero \cite{susyk}. (This is required by supersymmetry.)
The finite term in (\ref{sm2}) will thus be absent in such theories which are also known to be 
non-confining. This observation suggests a putative link between the finite terms in the EE and IR properties of  gauge theories. 
\end{enumerate}

As in the Abelian case, focusing on the gauge-invariant variable
$\phi^a$, we get only the part of EE, the nonabelian version of the
EE due to scalar fields, without the contact term.
 The wave function describes the vacuum properties
of the scalar field part, so this contribution may be referred to
as the contribution due to the wave function.
The latter term was due to the measure factor from gauge-fixing.
To see such an effect for the nonabelian theory, we must recast the formalism given here
in the language of gauge fixing.
As shown in \cite{KKN4}, this can be done.
 Notice that we can write
$A = M^{\dagger -1} (-\del H H^{-1}) M^\dagger + M^{\dagger -1} \del M^\dagger$, ${\bar A} = M^{\dagger -1} \bdel M^\dagger$, so that we may view the fields as a complex gauge transformation by $M^\dagger$ of the configuration $(A,  \bA) = (-\del H H^{-1}, 0)$.
The Gauss law condition on the wave functions
can then be used to eliminate $E$ which is conjugate to
$\bA$ in favor of $\bE$ in the expression for the Hamiltonian.
 This will involve some singular expressions which have to
 be evaluated with regularization and this leads to the mass term \cite{KKN4}.
 As far as the wave function is concerned, the physical variables one needs to
 take care of are $\bE$ and $A$. 
 The canonical one-form for the theory is given by
 \beq
 \A = \int E_i^a \delta A_i^a = - 4  \int \Tr ( \bE \, \delta A + E \, \delta \bA )
 \label{nonab5}
 \eeq
 where
 $E = (-i t^a) (E_1^a +i E^a_2)/2$,  $\bE = (-i t^a) (E_1^a - i E^a_2)/2$.
The generator of gauge transformations (or the Gauss law operator)
is
\beq
\G^a = 2 (\bD E + D \bE )^a
\label{nonab6}
\eeq 
We want to express $\A$ in terms of $\bE$, $A$, $\G$ and a conjugate constraint
which gives the required gauge choice, say, $\chi \approx 0$.
The gauge of interest can be viewed as $M^\dagger = 1$, but this is highly nonlocal in terms
of the original fields. What we need is a choice for which the commutator $[ \G(x), \chi (y)]$
is local, so that there is no additional source of entanglement. $\chi = \bA$ is possible, but here
the commutator is $\bD\, \delta (x-y)$ and it is not clear how we can split the chiral
operator $\bD$ into
contributions from two regions. So we choose $\chi = D \bA $.
The canonical one-form can then be written as
\beq
\A = - 4\int \Tr \Bigl[ \bE \, \delta A + \G (x)\,  (- D \bD )^{-1}_{x, y}\,  \delta \chi (y) \Bigr]
\label{nonab7}
\eeq
The phase volume then takes the form
\beq
d \mu =   \det [ (- D \bD )^{-1}]~ [d \bE d A ] \, [ d\G d\chi ]
\label{nonab8}
\eeq
or equivalently,
\beq
\det [ (- D \bD )] ~ d \mu =   ~ [d \bE d A ] \, [ d\G d\chi ]
\label{nonab9}
\eeq
The integral over $\G$, $\chi$ will have to be eliminated in the
functional integral via suitable integration. This could be over a $\delta$-function, or
over a contour enclosing $\chi = 0$ after a deformation of the $\chi$-contour into the
complex plane suitably. In any case,
we see that we get a factor $\det (- D \bD)_{\rm I}$ in region I,
$\det (- D \bD)_{\rm II}$ in region II, and a similar term for the full space.
Therefore, following the analysis for the Abelian case, we expect that the 
appropriate version of the contact term is given by
\beq
S_{\rm contact} = \log  \left[ { \det (- D \bD )_{{\rm I} \cup {\rm II} } \over \det (-D \bD )_{\rm I}
\, \det (- D \bD )_{\rm II} }\right]
\label{nonab10}
\eeq
This result depends on the fields, and so, in the expression for the entropy,
this will contribute with an averaging over the physical fields, i.e.,
the integration over $H$ has to be carried out.
We already have the mass term in this formulation, so we can consider
the expansion of (\ref{nonab10}) around the qualified {\it free} limit mentioned earlier.
The lowest order contribution from (\ref{nonab10}) is then the same as the result
for (dim$G$ copies of) the Abelian theory.
 
\section{The cone partition function and the contact term}
In this final section we connect our formulation of the contact term with the conventional results derived from the replica method.
The ground state wave functional at a fixed time, say at $t = 0$, can be obtained as the functional integral of $e^{- \S}$, where $\S$ is the Euclidean action, over all
fields for all $t < 0$ with specified fixed values at $t = 0$.
For the Maxwell field, we can use the BRST gauge-fixed Euclidean action
\beqar
\S &=& {1\over 4} \int F_{\mu\nu} \,F^{\mu\nu} ~+~ S_{gf}\nonumber\\
\S_{gf}&=&Q \int {\bar c} \,\left(  \del\cdot A - i {N \over 2}\right)
= \int i\,N\, \del\cdot A + {N^2 \over 2} + {\bar c} (- \square ) c
\label{BRST1}
\eeqar
Here $c$, ${\bar c}$ are the ghost fields, $N$ is the Nakanishi-Lautrup field.
The wave function may thus be written as
\beq
\Psi[\tilde A_\mu] = \int \mathcal{D}[A, c, \bar c] ~\delta (A_\mu(x_i, t = 0) - \tilde A_\mu (x_i)) ~e^{-S}\label{psi}
\eeq
If we apply  the replica trick directly to this expression
(\ref{psi}), the EE would still be given by (\ref{eff-s}), but $W_\alpha$ would now be given by \cite{kabat}
\beq
W_\alpha = \frac{1}{2}\tr\ln(g^{\mu \nu}(-\Box ) - R^{\mu \nu}) - \tr\ln(-\Box )\label{conem}
\eeq
The first term is the gauge field contribution, while the second term arises from the ghost fields. 
The functional determinants above are to be evaluated on the cone and the curvature term $R^{\mu \nu}$ represents a delta function contribution from the tip of the cone. Using the same techniques used in \cite{kabat}, one can obtain the following expression for the EE as derived from (\ref{conem}),
\beq
S^{cone}_E = (d-2)S_E(m=0) + S^{contact}_E
\label{cones}
\eeq
The last term in this equation, the so-called contact term, is given by
\beq
S^{contact}_E = -\frac{{\mathbb A}}{2\sqrt{\pi}}\left(\frac{1}{\e} + \mathcal{O}(\e)\right)
\label{contact}
\eeq
The expression for EE in (\ref{cones})
deviates from the expression obtained from the massless limit of (\ref{d3s}) due to the contact term contribution. To better understand this additional contribution it is useful to deconstruct the evaluation of (\ref{conem}), which is expressed in terms of the vector heat kernel
\beq
K_V(s,x,y)_{\mu \nu} = \sum_n e^{-\lambda_n s} A_{n \mu}(x) A_{n \nu}(y)
\eeq
Denoting the Lorentz indices along the cone (whose tip lies at the entangling surface) by $a,b$, the vector fields along those directions satisfy
\beq
(-g_{ab}\nabla^2 +R_{ab})A_n^b  = \lambda_nA_{na}
\eeq
These modes can be constructed from eigenfunctions of the scalar Laplacian $\phi_n$. Specifically, the  longitudinal and transverse components of $A_{a}$ are given by 
\beq
\frac{1}{\sqrt{\lambda _n}}\nabla_a \phi_n, \hskip .2in \frac{1}{\sqrt{\lambda_n}}\epsilon_{ab}\nabla^b\phi_n\label{conedual}
\eeq  
respectively \cite{kabat}. The direction transverse to the cone has no curvature contributions and the gauge field along that direction simply contributes one scalar degree of freedom to the partition function. After tracing over the $a,b$ indices, the heat kernel along the cone becomes \cite{kabat, DW}
\beq
K_V(s,x,x) = \sum_ne^{-\lambda _n s}\,\frac{1}{\lambda _n}\,2\, (\nabla _a\phi_n\nabla^a\phi_n) = 2K(s,x,x) + \underline{\int_s^\infty ds'\nabla^2K (s',x,x)}
\eeq
In the second expression, we have carried out an integration by parts (and $K(s)$ denotes the scalar heat kernel as before). The additional underlined term generates the contact contribution (\ref{contact}). Adding the scalar contribution from the direction transverse to the cone and those of the ghost fields, the above expression reproduces Kabat's result \cite{kabat} given in (\ref{cones}). We are now in a position to argue how the above contact term obtained from the conical partition function has the same physical origin as the one discussed earlier (\ref{glue6}). In our construction the contact term arose from unintegrated edge modes confined to the entangling surface. These are precisely the modes that the boundary term above captures which justifies our previous identification of $\det (M_{\rm I}
+ M_{\rm II})$ as the contact term.

\vskip .2in
We thank Daniel Kabat for many useful comments and discussions.
This research was supported in part by the U.S.\ National Science
Foundation grants PHY-1417562, PHY-1519449
and by PSC-CUNY awards.

\section*{Appendix A: Eliminating $\phi_0$ for the Maxwell-Chern-Simons theory}

Consider the MCS theory defined in a region, say I, with boundary.
The canonical one-form is given by
\beqar
\A&=& \int \left[ E_i \delta A_i - {m \over 2} \epsilon_{ij} A_i \delta A_j \right]\nonumber\\
&=&\int \left[ \del_i {\tilde \sigma} \del_i {\tilde \theta} + \del_i {\tilde \Pi} \del_i {\tilde \phi}
\right] + \A_{bndry} + \delta \int \left[ {m \over 2} \del_i \theta \del_i \phi \right]\label{A1}\\
\A_{bndry}&=& \oint \left[ \E + {m \over 2} \del_\tau \theta_0 \right] \, \delta \theta_0
+ \left[ Q - {m \over 2} \del_\tau \phi_0 \right]\, \delta \phi_0
\label{A2}
\eeqar
where $\E$ and $Q$ are, as in the Maxwell theory, given by
$\E (x) = \oint \sigma_0 (y) \, M(y,x) +\del_\tau \Pi_0 (x)$,
$Q(x) = \oint \Pi_0 (y) \, M(y,x) - \del_\tau \sigma_0 (x)$.
Because of (\ref{new4}) from the text, these still obey the constraint
\beq
\C = \del_x \oint_y \E (y) \, M^{-1}(y, x) + Q(x) \approx 0
\label{A3}
\eeq
The symplectic structure for the boundary fields is given by the boundary part
of $\A$ as
\beq
\Omega_{bndry} = \int \left[ \delta \E \, \delta \theta_0 + {m \over 2} \del_\tau \delta \theta_0 \, \delta \theta_0
+ \delta Q \, \delta \phi_0 - {m \over 2} \del_\tau \phi_0 \, \delta \phi_0
\right]
\label{A4}
\eeq
Using this, the Hamiltonian vector fields for the boundary fields are given by
\beqar
V_{\theta_0} &\longleftrightarrow& - {\delta \over \delta \E(x)} ,\hskip .2in
V_{\phi_0} \longleftrightarrow - {\delta \over \delta Q(x)}\nonumber\\
V_\E&\longleftrightarrow& {\delta \over \delta\theta_0(x)} + m 
{\del \over \del x} \left( {\delta \over \delta \E (x)}\right)\nonumber\\
V_Q &\longleftrightarrow& {\delta \over \delta\phi_0(x)} + m 
{\del \over \del x} \left( {\delta \over \delta Q (x)}\right)
\eeqar
with the Poisson brackets given by
$\{ F, G\} = - V_F \rfloor \delta G$. It is then easy to verify that $\{\C (x) , \C (y) \} = 0$, so that they remain first class even with the Chern-Simons term added to the action.
We can choose the conjugate constraint $\phi_0 \approx 0$ as before and eliminate it.
The canonical one-form thus reduces to
\beq
\A_{bndry} = \oint \left[ \sigma_0 (y) M(y, x) + \del_\tau \Pi_0 (x) 
+ {m \over 2} \del_\tau \theta_0 (x) \right] \, \delta \theta_0 (x)
\label{A6}
\eeq
This is what is used in text, see (\ref{new25}), (\ref{new25a}).

\section*{Appendix B: The topological contribution for the Maxwell-Chern-Simons theory}

The topological contribution in the case of pure Chern-Simons theory has been computed using numerous techniques in the literature \cite{CSE1,tope,CFTE, PIE}. We start with a brief
outline of the computation of the (topological) contribution to EE using the methods used in \cite{CSE1} which are closest in spirit to the Hamiltonian techniques employed in this paper. 

First of all, we make an
observation which establishes a point of contact with the papers cited
which use the Chern-Simons theory with a chiral field
on the boundary.
The Chern-Simons term is not invariant under gauge transformations which do not
vanish on the boundary. One can add a chiral field action on the boundary to
make a gauge-invariant action $\S_{\rm MCS} = \S_1 + \S_{ch}$, with
\beqar
\S_1 &=& \int d^3x~ \left[{1\over 2}( E^2 - B^2) + {m\over 2} \epsilon^{\mu\nu\alpha}
A_\mu \del_\nu A_\alpha \right] \nonumber \\
\S_{ch} &=&  {k e^2\over 4\pi} \int dt \oint \Bigl[ \del_0 \chi\, ( \del_{\tau} \chi + A_{\tau})
- \chi\, {\dot A}_{\tau} + A_0  \, A_{\tau}\Bigr] 
\label{16}
\eeqar
This is invariant under the gauge
transformation
$A_i \rightarrow A_i + \del_i f$ (or $\theta \rightarrow \theta + f$), 
$\chi \rightarrow \chi - f$, so that we may trade
the field $\chi$ for $\theta_0$ by choosing a gauge where $\chi$ is set to zero and
retaining $\theta_0$.
The resulting contribution to $\A$ is of the form
$ \del_\tau \theta_{0{\rm I}}
\delta \theta_{0{\rm I}} -  \del_\tau \theta_{0{\rm II}}
\delta \theta_{0{\rm II}}$ which is what occurs in (\ref{new31d}).
So we can use techniques similar to those for the chiral field in
\cite{CSE1}.

We consider the interface to be a circle of radius $R$,
coordinatized by 
$\tau$,  $0 \leq \tau \leq l$ with $l = 2 \pi R$.
Since $\theta$ is angle-valued field on the circle, it is a map
$\theta: S^1 \rightarrow S^1$.
Thus, in a general mode expansion for $\theta$, there is
a part which is
completely periodic
and a part which gives a shift  under $\tau \rightarrow \tau +l$.
 Since we have a $U(1)$ gauge symmetry,
it is sufficient for $e^{i \theta}$ to be periodic, so we can identify
$\theta$ and $\theta + 2 \pi {\mathbb Z}$, which shows that there
can be a nonzero shift $2 \pi {\mathbb Z}$.
The latter may be viewed as $\oint \del_\tau \theta$, or better as
a nontrivial holonomy around the circle
which can be accommodated by
a constant gauge connection $c$.

For the pure Chern-Simons action, 
we can drop the $\E$-dependent terms in 
(\ref{new31d}), (\ref{16}).
The canonical one-form for the Chern-Simons action
(or the corresponding part from the chiral action (\ref{16})), is then
\beq
\A =  {k \over 4\pi}  \oint   ( \del_{\tau} \theta + 2\,c)\, \delta \theta
\label{B1}
\eeq
We have added the constant flat connection $c$ to accommodate the
nonperiodicity of $\theta$.
(We have also absorbed $e^2$ into $\theta$, $c$. The new $\theta$ in 
(\ref{B1}) is periodic, $\theta (\tau +l ) = \theta (\tau )$.
The factor of $2$ for $c \delta\theta$-term is convenient for the following reason. 
The phase space function which
leads to the shift $\theta \rightarrow \theta + \epsilon$ via the Poisson brackets
defined by $\oint  \del_{\tau} \theta \delta \theta$ is $2 \del_\tau \theta$.
With the factor of $2$ for $c \delta \theta$, 
this function becomes $\del_\tau \theta + c$, which is a covariant derivative of
$\theta$ with connection $c$.)

The relevant terms in the action for the computation of the entanglement
entropy are then given by
\beqar
\S_{ch} &=& \frac{1}{4\pi}  \oint \left[ \partial_0 X^{\rm I}\partial_\tau X^{\rm I} -\partial_0 X^{\rm II}\partial_\tau X^{\rm II}  + 2\,C^{\rm I} \del_0 X^{\rm I}
- 2\,C^{\rm II} \del_0 X^{\rm II} \right]
- \int dt \, \H \label{period0}\\
\H&=& {v\over 4 \pi} \int  \Bigl[ (\partial_\tau X^{\rm I} + C^{\rm I})^2 + (\partial_\tau X^{\rm II}
+ C^{\rm II})^2 \Bigr] 
+ \frac{\lambda}{2\pi}\int\oint  \left[ 1 - \cos\frac{1}{\sqrt{k}}(X^{\rm I} - X^{\rm II})  \right] 
\nonumber
\eeqar
where we have written
$X^{\rm I} = \sqrt{k} \,\theta_{0{\rm I}}$, 
$C^{{\rm I}} = \sqrt{k} \, c_{\rm I}$, etc.
We have introduced the constraint term (\ref{new31i}) in the Hamiltonian
and also added an extra term for regularization.
The parameter $v$ can be regarded as a UV regulator which can eventually be set to zero. 
$\lambda \rightarrow \infty$ forces the fields to be identified on the entangling surface while preserving the periodicity constraints.

The chiral fields can be expanded in terms of their momentum modes as
\begin{eqnarray}
X^{\rm I} &=& X^{\rm I}_0 + \sum_{n<0}\left[ \frac{1}{\sqrt{|n|}}\alpha _n e^{2\pi i n\tau/l} +  \frac{1}{\sqrt{|n|}}\alpha^\dagger _n e^{-2\pi i n\tau /l}\right]\nonumber\\
X^{\rm II} &=& X^{\rm II}_0 + \sum_{n>0}\left[ \frac{1}{\sqrt{n}}\alpha _n e^{2\pi i n\tau/l} +  \frac{1}{\sqrt{n}}\alpha^\dagger _n e^{-2\pi i n\tau/l}\right]
\nonumber\\
C^{\rm I} &=& {2 \pi N^{\rm I}\over l} , \hskip .2in
C^{\rm II} = {2 \pi N^{\rm II}\over l}\,
\label{B1a}
\end{eqnarray}
It is easy to verify from the action (\ref{period0}) that the
``zero-mode operators" $X_0$ and $N$ are canonical conjugates, i.e.,
$[X_0^{\rm I}, N^{\rm I}] = i$, $[X_0^{\rm II}, N^{\rm II}] = -i$, as are the oscillator modes $[\alpha _n, \alpha ^\dagger _m] = \delta _{nm}$. 
In terms of the original $\theta$-variable, we have the
identification of $\theta$ with $\theta + 2 \pi {\mathbb Z}$.
With the redefined $\theta$, this implies that
the holonomy $\oint (\del_\tau \theta + c) = \oint c = 2 \pi {\mathbb Z}$. With the rescaling we have done,
this means that
$N^{{\rm I}/{\rm II}}$ are of the form $\sqrt{k} \,{\mathbb Z}$.
Further, for practical purposes, one expands the cosine above to quadratic order in 
$(X^{\rm I} - X^{\rm II})$.
The resultant Hamiltonian for the chiral modes can be expressed  as the sum of a zero-mode Hamiltonian
\beq
H_0 = \frac{\pi v}{2l}\left( (N^{\rm I} + N^{\rm II})^2 + l^2 ~\tilde{\lambda} (X_0^{\rm I} - X_0^{\rm II})^2\right)
\label{H0}
\eeq
and an oscillator Hamiltonian
\beq
H_\alpha = \frac{\pi v}{2l}\sum_{n \ne 0} \left( 4|n|\alpha ^\dagger _n\alpha _n  + 2 |n| + \frac{l^2 \tilde{\lambda} }{|n|}( \alpha _n \alpha ^\dagger _n -\alpha _n \alpha _{-n} + \alpha ^\dagger_n \alpha _n - \alpha ^\dagger _n \alpha ^\dagger _{-n} )\right) 
\label{Ha}
\eeq
where $\tilde{\lambda} = \lambda / (2 \pi^2 k v)$.
Further in deriving (\ref{H0}), (\ref{Ha}) we imposed the condition $(N^{\rm I} - N^{\rm II})\, |0\rangle = 0$ on the ground state. The zero-mode part has the form of a harmonic oscillator
and leads to a ground state wave function of the form
\beq
\vert \psi\ra = \sum_{n} \exp\left( - {(2 n\sqrt{k} )^2 \over 4 l {\tilde \lambda}}
\right) \, \vert N_{\rm I} \ra \otimes \vert N_{\rm II}\ra
\label{B2}
\eeq
where we use $n \sqrt{k} = N_{\rm I} = N_{\rm II}$. For the density matrix, the trace over 
the $\vert N_{\rm II}\ra$ states yields a reduced matrix
\beq
\rho = \sum_n \exp\left( - { 2\,(2 n\sqrt{k} )^2 \over 4 l {\tilde \lambda}}
\right)~ \vert N_{\rm I}\ra \la N_{\rm I}\vert
\label{B3}
\eeq
The exponent can be taken as the modular Hamiltonian for this case, and 
gives the partition function
\beq
Z_{zero} = \sum_n \exp\left( - {2 \beta l \over \sqrt{{\tilde \lambda}}} ( n /l)^2 k\right)
\approx \left({ l {\sqrt{\tilde \lambda}}\over 4 \beta}\right)^{\half}  \, \sqrt{{2\pi\over k}}
\label{B4}
\eeq
where we display the large $l$ behavior as the second approximate equality.

$H_\alpha $ can be diagonalized by a suitable Bogoliubov transformation.
The resulting density matrix leads to the partition function
\beq
Z_{osc} = \prod_{n>1} {1 \over \left( 1- e^{- 4 n \beta /l\sqrt{{\tilde \lambda}}} \right)}
\approx \left( {4 \beta \over l \sqrt{{\tilde\lambda}}}\right)^{{\half}}\,
\exp\left( {\pi^2 l \sqrt{{\tilde\lambda}}\over 24 \beta}\right)
\label{B5}
\eeq
The regularization-dependent prefactors cancel out in the product
giving the total partition function as
\beq
Z \approx \sqrt{{2\pi \over k}} \, \exp\left( {\pi^2 l \sqrt{{\tilde\lambda}}\over 24 \beta}\right)
\label{B6}
\eeq
This leads to the $- {\half} \log k$ in the entropy; this is the only $k$-dependence
in EE and is not dependent on the area of the entangling surface or the regularization.
Notice that this $k$-dependence is from the contribution of the zero modes.
The nonzero modes cancel some of the regularization-dependent terms.
The partition function (\ref{B6})
 leads to the entropy from the chiral boundary modes as \cite{CSE1}
\beq
S_{Chiral}(k) = \frac{\pi^2 }{12} {{\mathbb A}\over 2 \pi}
\sqrt{~\tilde{\lambda}} ~- ~\frac{1}{2}\log k + \cdots
\label{B7}
\eeq
where the ellipsis represent terms that are subleading in $1/l$. The first term is the cutoff dependent ``area'' term, with ${\mathbb A } = 2 \pi l$,
in which we see that the large $\lambda, l$ and small $v$ limits consistently reinforce each other. This term has the same structure as the leading divergent piece of the gauge field EE.
The  second term is the topological entropy.\footnote{This result can also be obtained using conformal field theory techniques \cite{CFTE} or by applying the replica trick to the Chern-Simons path integral \cite{PIE}.}

To apply this to the case of the Maxwell-Chern-Simons theory,
we start with the Green's function with Dirichlet boundary conditions
for the Laplacian on a disc. This is given by
\beq
G (r, \tau; r' , \tau'  ) =  {1\over 4\pi} \log \left[ {
R^2 ( r^2 + r'^2 - 2 r r' \cos (\vf -\vf')) \over
R^4 + r^2 r'^2 - 2 R^2 r r' \cos (\vf -\vf')}\right]
\label{B8}
\eeq
where $\vf = 2\pi \tau/l$. 
From this, we obtain
\beqar
M(\vf , \vf') &=& {1\over R^2}\left[ - {1\over 2\pi (1 - \cos (\vf -\vf' )}\right]\nonumber\\
&=& {1\over R^2}  \sum_{n=1}^\infty n \left[u_n (\vf )\, u_n (\vf') + v_n (\vf) \,v_n (\vf')\right]
\label{B9}\\
u_n (\vf) &=& {1\over  \sqrt{\pi}} \cos (n \vf ), 
\hskip .2in
v_n (\vf) = {1\over \sqrt{\pi}} \sin (n \vf ) \nonumber
\eeqar
$u_n$, $v_n$ are orthonormal mode functions (with integration over $\vf$ rather than $\tau$).
Thus, apart from the $R^{-2}$ factor which can be absorbed into integration variables,
\beq
M^{-1} =  \sum_{n=1}^\infty {1\over n}  \left[u_n (\vf )\, u_n (\vf') + v_n (\vf) \,v_n (\vf')\right]
\label{B10}
\eeq
It is easy to verify that $\del_\vf \del_{\vf'} M^{-1}= M$ as in (\ref{new4}).
Consider now the canonical one-form for the MCS theory given in
(\ref{new31d}), with the addition of the flat connection $c$; i.e.,
\beqar
\A (\alpha_0, \theta_0 ) &=&
\oint \left[\E_{\rm I}\, \delta \theta_{0{\rm I}} + e^2 \left\{ {k \over 4\pi }\, \del_\tau \theta_{0{\rm I}}
\delta \theta_{0{\rm I}}  + {k \over 2 \pi}  c_{\rm I} \delta \theta_{0{\rm I}}\right\} \right] ~+\nonumber\\
&&\hskip .2in
\oint \left[ \E_{\rm II} \, \delta \theta_{0{\rm II}} - e^2 \left\{{k \over 4\pi } \,\del_\tau \theta_{0{\rm II}}
\delta \theta_{0{\rm II}} +{k  \over 2 \pi}  c_{\rm II} \delta \theta_{0{\rm II}}\right\} \right]
\label{B11}
\eeqar
with $\E_{{\rm I}/{\rm II}} =  \alpha_{0{{\rm I}/{\rm II}} }\, M_{{\rm I} /{\rm II}} \pm \del_\tau \Pi_{0{{\rm I}/{\rm II}}}$
With the mode expansion (\ref{B1a}), we can verify that the terms
$\oint [\alpha_0 (\vf) M(\vf, \vf') \pm \del_\tau \Pi_0 (\vf' ) ] \delta \theta_0 (\vf')$
do not have a contribution from $X_0^{{\rm I}/{\rm II}}$.
The ``zero mode" fields $X_0^{{\rm I}/{\rm II}}$, $N^{{\rm I}/{\rm II}}$
are decoupled from the nonzero modes in the expression for $\A$.
For the nonzero modes, we have the straightforward identification
of $X^{\rm I}$ with $X^{\rm II}$.
 Therefore, the cancellations for the terms involving
$\del_\tau \Pi_0 \delta \theta$, $\del_\tau \theta \, \delta \theta$ between
I and II as mentioned
in text (see (\ref{new25a})) apply and we can simplify $\A$ to
\beq
\A = \oint \alpha_{0{\rm I}} M_{\rm I}\, \delta \theta_{0{\rm I}} + \oint \alpha_{0{\rm II}} M_{\rm II}\, \delta \theta_{0{\rm II}} + N^{\rm I} d X_0^{\rm I} - N^{\rm II} dX_0^{\rm II}
\label{B12}
\eeq
The analysis of the zero mode part proceeds as in the Chern-Simons case, and
we obtain the same topological contribution to the entropy.
With the constraints $\alpha_{0{\rm I}} - \alpha_{0{\rm II}} \approx 0$,
$\theta_{0{\rm I}} - \theta_{0{\rm II}} \approx 0$, we recover the arguments
given in text in section 4, leading to the contact term as in
(\ref{new31a}).


\end{document}